\documentstyle[preprint,prb,aps]{revtex}
\begin{document}

\draft

\title{Numerical Study of a Two-Dimensional Quantum
       Antiferromagnet with Random Ferromagnetic Bonds}

\author{Anders W. Sandvik\cite{aboakademi}}
\address{Department of Physics, University of California,
Santa Barbara, CA 93106}

\date{July 21, 1994}

\maketitle

\narrowtext

\begin{abstract}
A Monte Carlo method for finite-temperature studies of the two-dimensional
quantum Heisenberg antiferromagnet with random ferromagnetic bonds is
presented. The scheme is based on an approximation which allows for an
analytic summation over the realizations of the randomness, thereby
significantly  alleviating the ``sign problem'' for this frustrated spin
system. The approximation is shown to be very accurate for ferromagnetic bond
concentrations of up to ten percent. The effects of a low concentration of
ferromagnetic bonds on the antiferromagnetism are discussed.
\end{abstract}
\vskip0.5cm


\vfill\eject

Monte Carlo studies of frustrated quantum spin systems are difficult since
positive definite weight functions cannot be constructed in general
[the so called ``sign problem''].\cite{miyashita,loh}  For random models, on
the other hand, averaging over a large number of realizations of the
randomness is necessary, which considerably increases the computational
effort over what is required for non-random systems. Both the above
difficulties are present for the antiferromagnetic Heisenberg model with
random ferromagnetic bonds. The two-dimensional (2D) version of this model
is of current interest as a possible model of the copper-oxygen sheets of
lightly doped, but still insulating high-$T_c$ superconductor materials.
The idea, stressed by Aharony {\it et al.},\cite{aharony} is that in the
doped insulating phase, the holes introduced into the Cu-O sheets are
localized at individual oxygen sites. The coupling between the copper and
oxygen spins results in an effective {\it ferromagnetic} coupling
between the copper spins adjacent to an oxygen spin. Due to the
computational problems mentioned above, this picture has not yet
been tested by direct numerical calculations of experimentally measurable
quantities of the proposed model hamiltonian.

Previous numerical work on random quantum spin systems has been largely
limited to 1D systems\cite{random1d} and non-frustrated 2D models.
\cite{random2d} Quantum Monte Carlo simulations of random systems with
frustration have been carried out in cases where the sign problem is not
present, such as the Ising spin glass in a transverse field.\cite{ising}
For models with random long-range interactions, recent progress has been
made using field-theoretic methods.\cite{sachdev}

In this paper a Monte Carlo method for finite-temperature studies of
the 2D Heisenberg model with mixed antiferromagnetic and ferromagnetic
nearest-neighbor couplings of equal strengths is presented. The scheme
employs an approximation which corresponds to an annealing of the quenched
disorder. This approximation is argued to be very accurate in the regime
of interest for the high-T$_c$ cuprates; a concentration of ferromagnetic
bonds of less than ten percent. The summation over all realizations of the
annealed randomness can be carried out analytically for each Monte Carlo
configuration, thereby significantly alleviating the sign problem.
Furthermore, in a single simulation, calculations can be carried out for
several concentrations of ferromagnetic bonds with essentially no additional
computational cost.

Below, the method is described and tested for small systems. Results are
presented for the effect of an increasing concentration of ferromagnetic
 bonds on the staggered structure factor and the uniform susceptibility.

The model is defined by the hamiltonian
\begin{equation}
\hat H = \sum\limits_{\langle i,j\rangle}J_{ij} \vec S_i \cdot \vec S_j,
\label{hamiltonian}
\end{equation}
where ${\langle i,j\rangle}$ is a pair of nearest-neighbor sites on a
square lattice, and $\vec S_i$ is a spin-$1\over 2$ operator at site $i$.
The coupling constants $J_{ij}$ are all of equal strength $J$, but their
signs are random, with a probability $\rho$ for $-J$ (ferromagnetic)
and $1-\rho$ for $+J$ (antiferromagnetic). In its current formulation
the method to be presented does not allow for different ferromagnetic and
antiferromagnetic coupling strengths. The random $\pm J$ model should, however,
exhibit the general features associated with the presence of a low
concentration of frustrating bonds.

The computational scheme will be discussed in the context of a
generalization of Handscomb's quantum Monte Carlo method, \cite{sandvik}
but the same idea should be applicable to ``world-line'' methods
\cite{worldline} as well. Consider the expectation value of an
operator $\hat A$ at inverse temperature $\beta=1/{k_BT}$:
\begin{equation}
\langle \hat A \rangle = {1\over Z} \hbox{Tr} \lbrace \hat A
\hbox{e}^{-\beta \hat H} \rbrace ,\qquad
Z=\hbox{Tr}\lbrace \hbox{e}^{-\beta \hat H} \rbrace .
\label{expa}
\end{equation}
The starting point for the generalization of Handscomb's method is to
Taylor expand e$^{-\beta\hat H}$ and to write the traces in
(\ref{expa}) as sums over diagonal matrix elements in a suitably chosen
basis $\lbrace |\alpha \rangle \rbrace$, giving for the partition
function
\begin{equation}
Z = \sum\limits_{n=0}^\infty \sum\limits_{\alpha} {(-\beta)^n\over n!}
\langle  \alpha | \hat H^n | \alpha \rangle .
\label{part1}
\end{equation}
For the Heisenberg model, the basis
$\lbrace | S^z_1, \ldots , S^z_N \rangle \rbrace$, $S^z_i \in \lbrace
\uparrow ,\downarrow\rbrace$, is chosen, and the hamiltonian
(\ref{hamiltonian}) is written as
\begin{equation}
\hat H = -{J\over 2}\sum\limits_{b=1}^{N_b} \sigma_b \Bigl [ H_{1,b} -
H_{2,b} \Bigr ] + {N_b(1-2\rho)J/4},
\label{ham2}
\end{equation}
where
\begin{eqnarray}
\hat H_{1,b} = && ~2(\hbox{$1\over 4$}-S^z_{s1(b)}S^z_{s2(b)}) \nonumber \\
\hat H_{2,b} = && ~S^+_{s1(b)}S^-_{s2(b)} + S^-_{s1(b)}S^+_{s2(b)} .
\end{eqnarray}
Here $s1(b)$ and $s2(b)$ are the sites connected by bond $b$, $N_b=2L^2$ is
the number of bonds of the lattice, and $\sigma_b$ is $-1$ if $b$ is a
ferromagnetic bond and $+1$ otherwise. The partition function can now be
written as
\begin{equation}
Z = \sum\limits_{n=0}^\infty \sum\limits_{S_n}
\sum\limits_{\alpha} {(-1)^{n_F} ({\beta J / 2} )^{n} \over n!}
\Bigl\langle\alpha\Bigl | \prod\limits_{i=1}^n H_{a_i,b_i} \Bigr |
\alpha \Bigr \rangle ,
\label{part2}
\end{equation}
where $S_n$ denotes a sequence of $n$ index pairs,
\begin{equation}
S_n = {a_1\choose b_1}_1{a_2\choose b_2}_2 \ldots {a_n\choose b_n}_n,
\end{equation}
with $a_i \in \lbrace 1,2\rbrace $, $b_i \in \lbrace 1,\ldots ,N_b\rbrace$
referring to an operator $\hat H_{a,b}$. The matrix element
in (\ref{part2}) is equal to $0$ or $1$, and the sign of a given term
is determined only by the number $n_F$ of operators $\hat H_{a,b}$ with $b$
being one of the ferromagnetic bonds. This sign rule is valid for a bipartite
lattice, in which case an operator string contributing to $Z$ must flip
each spin in $|\alpha \rangle$ an even number of times, and therefore the
total number of operators $\hat H_{2,b}$ must be even. Note that the only
dependence on the realization of the randomness in (\ref{part2}) is in the
number $n_F$. This is crusial in what follows.

The actual Monte Carlo scheme has been described elsewhere, \cite{sandvik}
and will not be discussed here. It suffices to note that, as has been
shown above, for a
given realization $R$ of the $\pm J$ bonds on the lattice, an operator
expectation value can be written as
\begin{equation}
\langle \hat A\rangle_R = {\sum\limits_{C} W_CA_CS_C(R) \over
\sum\limits_{C} W_CS_C(R) },
\label{rexp}
\end{equation}
where $W_C$ is a positive definite weight for the configuration
$C$ [$C$ here belongs to the space
$\lbrace \alpha ,S_n, n=0,1,2,\ldots \rbrace$ of
states and index sequences], $S_C(R)=(-1)^{n_F(R)}$ is a sign which depends on
the realization $R$ as well as $C$, and $A_C$ is a function measuring the
operator $\hat A$ [the construction of $A_C$ for various types of operators
is discussed in detail in Ref. \onlinecite{sandvik}].
Here only operators without
explicit dependence on the particular realization of the randomness will be
considered, e.g. bulk susceptibilities and magnetic structure factors.

In a Monte Carlo simulation the configurations $C$ are generated using
$W_C$ as a relative probability distribution, and the quantities $S_C(R)$
and $A_CS_C(R)$ are measured with regular intervals. The expectation value
of $\hat A$ is then given by\cite{worldline}
\begin{equation}
\langle \hat A\rangle_R = {\langle A_CS_C(R) \rangle \over
\langle S_C(R) \rangle } ,
\label{aver1}
\end{equation}
and the average over the realizations of the randomness is
\begin{equation}
\langle\langle \hat A\rangle\rangle = {1\over N_R} \sum\limits_R
{\langle A_CS_C(R) \rangle \over
\langle S_C(R) \rangle} ,
\label{aver2}
\end{equation}
where $N_R$ is the number of realizations. If $\langle S_C(R) \rangle \ll 1$,
accurate determinations of $\langle S_C(R) \rangle$ and
$\langle A_CS_C(R) \rangle$ become very time consuming. Since
$\langle S_C(R) \rangle$ in most cases approaches zero exponentially as
the temperature is lowered, the sign problem is a severe limitation of
the quantum Monte Carlo technique for models where one cannot
construct [e.g. using symmetries] a weight function with a sign
identically equal to one.\cite{loh}

Since the weight $W_C$ does not depend on the realization of the
randomness, an estimate of $\langle\langle \hat A\rangle\rangle$ can be
obtained by carrying out the measurements of $A_CS_C(R)$ and $S_C(R)$ on a
set of pre-generated realizations in a single Monte Carlo simulation.
Hence, the randomness averaging does not pose a problem. However, this by
itself does not alleviate the sign problem, as the evaluation of the
individual terms of (\ref{aver2}) still becomes unstable when $S_C(R) \ll 1$.
The approximation introduced next will be shown to significantly reduce
this sign problem.

Consider the expectation values $\langle A_CS_C(R) \rangle$ and
$\langle S_C(R) \rangle$ averaged over the randomness:
\begin{eqnarray}
\langle \langle S_C \rangle\rangle~ && = {1\over N_R} \sum_R
\langle S_C(R) \rangle \nonumber \\
\langle \langle A_CS_C \rangle\rangle~ && = {1\over N_R} \sum_R
\langle A_CS_C(R) \rangle
\end{eqnarray}
The realization-dependent averages  can be written in terms of their
deviations from the respective realization-averaged quantities as
\begin{eqnarray}
\langle A_CS_C(R) \rangle~ && =
\langle \langle A_CS_C \rangle\rangle + \Delta_{AS}(R) \nonumber \\
\langle S_C(R) \rangle~ && =
\langle \langle S_C \rangle\rangle + \Delta_{S}(R) .
\end{eqnarray}
A realization-averaged expectation value can then be written as
\begin{equation}
\langle\langle \hat A\rangle\rangle =
{\langle \langle A_CS_C \rangle\rangle \over
\langle \langle S_C \rangle\rangle } + {\cal O}(\Delta^2),
\label{approx}
\end{equation}
where ${\cal O}(\Delta^2)$ denotes terms of order
$\Delta_{AS}(R)\Delta_{S}(R)$ and $(\Delta_{S}(R))^2$. [Note that
(\ref{approx}) would be exact if $\Delta_S(R)$ would be zero for
all $R$, even with $\Delta_{AS}(R) \not=0$.] If the concentration
of ferromagnetic bonds is low, the approximation (\ref{approx}) can be
expected to be a good one, since the main contribution to
$\langle\langle \hat A\rangle\rangle$ is from realizations where the
ferromagnetic bonds are far apart from each other. The signs
$\langle S_C(R) \rangle$ should then typically be insensitive to variations
in $R$. Note, however, that the approximation {\it does} contain collective
impurity effects, as the averages in (\ref{approx}) depend in a non-trivial
manner on the number of ferromagnetic bonds present.

Under the above approximation, the averages over the Monte Carlo
configurations and the realizations of the randomness have been put
on an equal footing. This corresponds to going from quenched to
annealed disorder. The computational advantage is that the
averaging over the randomness can be performed {\it analytically}
for each Monte Carlo configuration, which in effect means that
each Monte Carlo measurement step corresponds to measuring on a very large
number $N_R$ of configurations, which for a fixed number $N_f$ of
ferromagnetic bonds is given by
\begin{equation}
N_R = {N_b \choose N_f} .
\end{equation}
One might hope that this averaging enables a stable evaluation of the
expectation values $\langle \langle S_C \rangle\rangle$ and
$\langle \langle A_CS_C \rangle\rangle$ far beyond the point where estimates
of $\langle A_CS_C(R) \rangle$ and $\langle S_C(R) \rangle$ become too noisy.

Denoting by $F(R)$ the set of ferromagnetic bonds in the realization
$R$, the sign of a Monte Carlo configuration can
be written as
\begin{equation}
S_C(R)  = \prod\limits_{b\in F(R)} (-1)^{n_b} =
\prod\limits_{b\in F(R)} s_C(b),
\end{equation}
where $n_b$ is the number of operators acting on bond $b$, i.e. the
number of index pairs $a_i \choose b_i$ with $b_i=b$ in $S_n$.
Hence the sign is a product of ``local signs'' $s_C(b)$, with
$s_C(b)$ being positive or negative depending on if $n_b$ is even or odd.
Note that the local signs depend only on the Monte Carlo configuration
$C$, and the full sign $S_C(R)$ is calculated using only the local signs of
the ferromagnetic bonds of $R$. Denoting the total number of local minus
signs by $n_-$ and the total number of local plus signs by $n_+=N_b-n_-$,
the randomness
averaged sign $\Sigma_C={1\over N_R}\sum_R S_C(R)$ of a Monte Carlo
configuration is given by
\begin{equation}
\Sigma_C = {1\over N_R} \sum\limits_{f=0}^{N_f} (-1)^f
{n_- \choose f}{n_+\choose N_f -f} .
\label{sigmac}
\end{equation}
The (approximate) Monte Carlo estimate (\ref{approx}) for the disorder
averaged $\langle \hat A\rangle$ can now be written as
\begin{equation}
\langle\langle \hat A\rangle\rangle = {\langle A_C\Sigma_C \rangle \over
\langle \Sigma_C \rangle},
\label{aver3}
\end{equation}
where all effects of the randomness is contained in $\Sigma_C$, which
can be easily calculated for each Monte Carlo configuration.\cite{sign}

Next it will be demonstrated that this estimate of
$\langle\langle \hat A\rangle\rangle$ is indeed considerably less noisy than
(\ref{approx}), and that the approximation involved is very good, at least
when the concentration of ferromagnetic bonds is low.
Results will be shown for the staggered structure factor
\begin{equation}
S(\pi,\pi) = {1\over L^2}\sum\limits_{j,k}
\hbox{e}^{i\vec\pi \cdot (\vec r_k-\vec r_j)}
\langle\langle S^z_jS^z_k \rangle\rangle
\end{equation}
and the uniform susceptibility
\begin{equation}
\chi(0,0) =  {1\over L^2}\sum\limits_{j,k} \int\limits_0^\beta d\tau
\langle\langle S^z_j(\tau)S^z_k(0) \rangle\rangle .
\end{equation}
In order to test the accuracy of the ``annealed'' approximation,
simulations of $L\times L$ systems with $L=4$ and $8$ were carried out,
and $S(\pi,\pi)$ and $\chi(0,0)$ were calculated using both
(\ref{aver2}) and (\ref{aver3}) [With (\ref{aver2}), the averaging over $R$
was done for using several hundred randomly generated realizations]. Fig. 1
shows results for $L=4$ at temperatures $T/J =0.4,0.6$, and $0.8$.
The maximum ferromagnetic bond concentration for which the averages
can be evaluated decreases rapidly as the temperature is lowered.
As expected, the approximate averages are easier to obtain than the
exact ones. Perhaps surprisingly, no deviations of the approximate
averages from the exact ones can be seen within statistical errors, even
for rather high $\rho$. The antiferromagnetism is strongly suppressed by
the disorder; the staggered structure factor decreases with $\rho$ and
the uniform susceptibility is enhanced. The effect becomes stronger as
the temperature is decreased. Fig. 2 shows similar results for $8\times 8$
systems. Here the suppression of the antiferromagnetism is even stronger.
Again, no differences between the approximate and exact results can be seen
up to the maximum $\rho$ for which they can both be relyably calculated.
One might expect the errors of the annealed approximation to become
larger at lower temperatures, where comparisons are difficult due to the
sign problem.

Fig. 3 shows the average sign versus the
number of ferromagnetic bonds for $L=4$ and $L=8$ at $T/J = 0.4,0.6$,
and $0.8$. There is very little size-dependence, confirming that the
average sign at a given temperature depends essentially only on
the number of ferromagnetic bonds present. Note that $\langle \Sigma_C\rangle$
can be accurately evaluated for the larger system even when it becomes
extremely small.\cite{sign} Using the exact expression (\ref{aver2}) is not
feasible if $\langle S_C(R) \rangle$ becomes smaller than
$\approx 10^{-3}$. This limits the maximum number of ferromagnetic
bonds that can be studied. An accurate calculation of
$\langle \Sigma_C\rangle$, on the other hand, is possible up to some
maximum $\rho$, which is essentially independent of the system size.
The evaluation of the expression (\ref{aver3})
still becomes more difficult as the system size increases for operators such as
the staggered structure factor, for which the autocorrelation time
grows with the system size and fluctuations in $\langle A_C\Sigma_C\rangle$
become problematic.

The model (\ref{hamiltonian}) has antiferromagnetic long-range order
at $T=0$ for $\rho \to 0$ and ferromagnetic order for $\rho \to 1$.
At intermediate $\rho$ there is presumably a spin-glass phase. An important
open question is the critical concentration of ferromagnetic bonds
needed to destroy the antiferromagnetism. In principle finite-size scaling
of the staggered structure factor can answer this question, which however
is beyond the scope of this paper. Here some initial results for the
effect of an increasing fraction of ferromagnetic bonds on systems
of size $10\times 10$ are presented.

In Fig. 4 the staggered structure factor is graphed versus
the temperature for various concentrations of ferromagnetic bonds. For
$\rho = 2.5\%$ and $5\%$, $S(\pi,\pi)$ is significantly suppressed, but still
has a temperature dependence similar to the clean system. For $\rho = 10\%$
the structure factor becomes almost temperature-independent at
$T \approx J/2$. This might be an indication that no long-range order
exists for this concentration.

Fig. 5 shows the enhancement of the uniform susceptibility as the
disorder is increased. For comparison, $\rho=0$ results for $L=64$
are also shown. The finite-size effects for the uniform susceptibility
are apparently quite small. The enhancement due to the presence of
ferromagnetic bonds is significant already for $\rho = 2.5\%$. As $\rho$
is increased one would expect $\chi(0,0)$ to eventually diverge as $T\to 0$.
There are indications of such behavior for $\rho \ge 5\%$, but unfortunately
the sign problem limits the accuracy in this regime.

It is unclear whether low enough temperatures can be reached for
determining the critical concentration using the scheme presented here.
It should be possible, however, to carry out detailed studies for
larger systems at temperatures above $T \approx J/4$ for concentrations
of a few percent. This should enable an assessment of the relevance of
the model to the high-T$_c$ cuprates. The high-temperature regime is also
important in view of the recent work on 2D quantum antiferromagnetism
based on the nonlinear sigma-model.\cite{sigmamodel}

In summary, a scheme which alleviates the sign problem in
quantum Monte Carlo simulations of the 2D quantum Heisenberg antiferromagnet
with random ferromagnetic bonds has been presented. Results for the
staggered structure factor and the uniform susceptibility show that the
presence of a few percent of ferromagnetic bonds substantially suppresses
the antiferromagnetism. The method discussed here can easily be extended
for 3D systems.

I would like to thank M. Boninsegni, D. Scalapino, and M. Veki\'c for
stimulating discussions. This work is supported by the Department of
Energy under Grant No. DE-FG03-85ER45197.

\begin{figure}
FIG. 1.
The staggered structure factor and the uniform susceptibility versus
the ferromagnetic bond concentration for $L=4$ at three different
temperatures. Results obtained using (\ref{aver2}) are shown
as solid squares ($T/J=0.8$), open circles ($T/J=0.6$), and
solid circles ($T/J=0.4$). The solid curves are drawn through points
obtained using the ``annealed'' approximation ({\ref{aver3}}).
\end{figure}

\begin{figure}
FIG. 2.
Same as Fig. 1 for systems of size $8\times 8$.
\end{figure}

\begin{figure}
FIG. 3.
The average sign versus the number of
ferromagnetic bonds for $L=4$ (open symbols) and $L=8$ (solid symbols).
Circles are for $T/J=0.8$, squares for $T/J=0.6$, and triangles for
$T/J=0.4$.
\end{figure}

\begin{figure}
FIG. 4.
The staggered structure factor for $L=10$ versus the temperature for
$\rho =0\%, 2.5\%, 5\%, 8\%$, and $10 \%$ ($S$ decreasing with $\rho$).
\end{figure}

\begin{figure}
FIG. 5.
The uniform susceptibility for $L=10$ versus the temperature for
$\rho =0\%, 2.5\%, 5\%, 8\%$, and $10 \%$ ($\chi$ increasing with $\rho$).
The dashed curve goes through $\rho=0$ results calculated for a system
of size $64\times 64$.

\end{figure}


\begin{references}
\bibitem{sign}
For large systems, care has to be taken in the evaluation of the
average sign, as $\Sigma_C$ can be very small relative to the
individual terms of Eq. (\ref{sigmac}), and therefore highly
sensitive to truncation errors.
\end{references}
\end{document}